\documentclass{article}

\usepackage{arxiv}
\usepackage[utf8]{inputenc} 
\usepackage[T1]{fontenc}    
\usepackage{hyperref}       
\usepackage{url}            
\usepackage{booktabs}       
\usepackage{amsfonts}       
\usepackage{nicefrac}       
\usepackage{microtype}      
\usepackage{lipsum}
\usepackage{graphicx}
\usepackage{color}
\usepackage{amssymb}
\usepackage{amssymb}
\usepackage{amsmath}
\usepackage{mathtools}
\usepackage{geometry}
\usepackage{pdflscape}
\usepackage{rotating}
\usepackage{setspace}

\usepackage[table,xcdraw]{xcolor}

\newcommand{\avg}[1]{\langle{#1}\rangle}
\newcommand{\eigvec}[2]{\mathbf{v}^\mathbf{#1}_\i{#2}}
\newcommand{\eigval}[2]{\lambda^\mathbf{#1}_\mathrm{#2}}
\newcommand{\scal}[2]{{#1} \cdot {#2}}

\newcommand{\Cemp}{\mathbf{C}}
\newcommand{\Dmat}{\mathbf{D}}

\renewcommand{\i}[1]{\mathrm{{#1}}}

\title{Conditional Correlations and Principal\\Regression Analysis for Futures}

\author{
  Armine Karami \\%
 Chair of Econophysics \& Complex Systems,\\
  LadHyX, Ecole polytechnique, 91128 Palaiseau\\
  \texttt{armine.karami@ladhyx.polytechnique.fr} \\
   \And
  Raphael Benichou \\
  Capital Fund Management,\\
  23-25 Rue de l'Universit\'e, 75007 Paris \\
  \texttt{raphael.benichou@cfm.fr}
   \AND
  Michael Benzaquen \\
  Chair of Econophysics \& Complex Systems,\\
 LadHyX, Ecole polytechnique, 91128 Palaiseau\\
  \texttt{michael.benzaquen@polytechnique.edu} \\
  \And
  Jean-Philippe Bouchaud \\
  Capital Fund Management,\\
  23-25 Rue de l'Universit\'e, 75007 Paris \\
  \texttt{jean-philippe.bouchaud@cfm.fr}
}

\begin{document}
    \maketitle


\onehalfspacing

    \begin{abstract}
        We explore the effect of past market movements on the instantaneous correlations between assets within the futures market. Quantifying this effect is of interest  to estimate and manage the risk associated to portfolios of futures in a non-stationary context. We apply and extend a previously reported method called the Principal Regression Analysis (PRA) to a universe of $84$ futures contracts between $2009$ and $2019$. We show that the past up (resp. down) 10 day trends of a novel predictor -- the eigen-factor -- tend to reduce (resp. increase)  instantaneous correlations. We then carry out a multifactor PRA on sectorial predictors corresponding to the four futures sectors (indexes, commodities, bonds and currencies), and show that the effect of past market movements on the future variations of the instantaneous correlations can be decomposed into two significant components. The first component is due to the market movements within the index sector, while the second component is due to the market movements within the bonds sector. 
    \end{abstract}


    \section{Introduction}

        A crucial input for managing portfolio risk is the covariance matrix of the underlying assets. The empirical determination of this matrix is fraught with difficulties. One is that of sample size: when the length of the available time series is not very large compared to the number of assets in the portfolio, the empirical covariance matrix suffers from very significant biases. For example, the smallest eigenvalue is underestimated and the largest eigenvalue overestimated. The corresponding eigenvectors are also strongly affected by measurement noise. As a consequence, the realized risk of optimal (Markowitz-like) portfolios can be considerably larger than anticipated, see e.g. \cite{bun2017reports}.  
        
        But there is another, perhaps more fundamental reason for the out-of-sample risk to be larger than expected: we do not live in a stationary world, described by a time-invariant covariance matrix. The covariance between assets evolves not only because the volatility of each asset changes over time \cite{cont_stylized} and react to the recent market trend \cite{bekaert2000asymmetric,matacz_leverage, li2005relationship}, but also because the correlations themselves increase or decrease, depending on market conditions --- see e.g. \cite{Ang2002,borland2012statistical,balogh2010persistent}. Sometimes these correlations jump quite suddenly, due to an unpredictable geopolitical event. The arch example of such a scenario is the Asian crisis in the fall of 1997, when the correlation between bonds and stocks indexes abruptly changes sign and become negative --- a ``flight to quality'' mode that has prevailed ever since \cite{wyart,baur2009flights}. Whereas these events are hard to predict, some measurable indicators (or factors) do anticipate future changes of the  correlation matrix. For example, as documented in \cite{reigneron2011principal}, the structure of the stock correlation matrix depends (statistically) on the past returns of the stock index (i.e. the average of all stock returns). The main effect is that the top eigenvalue of the correlation matrix increases after the whole market goes down (and vice versa). Correspondingly, the top eigenvector rotates towards  $\mathbf{u}_0=(1,1,\cdots,1)/\sqrt{N}$ after a drop of the index.  
        
        The main idea of Ref. \cite{reigneron2011principal} is to regress the instantaneous correlation matrix on the past value of one or several indicators $I_\i{a}$, as follows:
        \[
        r_\i{i}(t) r_\i{j}(t) = \Cemp_{\i{ij}} + \sum_\i{a} \Dmat^\i{a}_\i{ij}(\tau) I_\i{a}(t-\tau) + \mathbf{\varepsilon}_\i{ij}(t, \tau)\, ,
        \]
        where $r_\i{i}(t)$ is the standardized return of asset $i$ between $t$ and $t+1$,  $\mathbf{\varepsilon}$ a zero-mean noise term and $I_\i{a}(t-\tau)$ the value of the indicator `a' fully known at time $t$. This indicator is assumed to be of zero mean, such that $\Cemp$ is by definition the unconditional correlation matrix. Finally, $\Dmat^{\text a}(\tau)$ is a matrix that measures the sensitivity of the instantaneous correlation matrix to indicator $I_\i{a}$ lagged by $\tau$. Low rank approximations of the $\Dmat^{\text a}(\tau)$ matrices allow one to build intuitive and parsimonious models of how correlations are affected by the past. The eigenvalues and eigenvectors of this matrix are expected to describe the impact of $I_\i{a}(t - \tau)$ on correlations in a more precise way than what one would obtain by separate examination of the slope coefficients $\Dmat_\i{i,j}(\tau)$. This protocol has been dubbed Principal Regression Analysis (PRA) in \cite{reigneron2011principal}, and has the virtue of being much easier to calibrate and to interpret than Dynamical Conditional Correlation (DCC) models, see for example \cite{Engle2002}.
        
        The aim of the present study is to extend to futures market the analysis of \cite{reigneron2011principal}, which was devoted to individual stocks. This is interesting for different reasons. One is that the CTA (Commodity Trading Advisor) industry routinely deals with portfolios of futures, and the risk of these portfolios is obviously an important aspect CTA funds want to monitor and control. Second, most  stocks are positively correlated and, correspondingly, the top eigenvector of the correlation matrix always remains close to the uniform mode $\mathbf{u}_0$. This is not the case in the universe of futures contracts. For example, as mentioned above, stock indexes and bonds are typically negatively correlated (at least since 1998). 
        
        In this paper, we apply a Principal Regression Analysis to the universe of futures contracts, and elicit the main factors affecting the structure of the corresponding correlation matrix. Whereas the dynamical structure of the correlation matrix  of stock returns has been discussed in several studies, using different methods, we are not aware of similar investigations of futures returns.  We first analyse the simplest case of a single  factor, that we choose to be a `hand-made' index $I_0$, where stock indexes, currencies (vs. dollar) and commodities all have the same positive weight, and bonds have a negative weight of equal magnitude. We study the time scale over which the return of this index should be measured such that the effect on the correlation matrix is strongest, and find  $\approx 10$ days. We then replicate the analysis with  an index constructed from the top eigenvector of $\Cemp$ which is close to, but not identical to $I_0$, and find that the quality of  the prediction is increased. Finally, we run a multivariate PRA using four factors, namely the past returns of the four relevant sectors in the universe of futures: stock indexes, bonds, currencies and commodities. This allows us to get a more complete picture of the  mechanisms leading to a change in the correlation structure of futures markets. 


    \section{Data and notations}
    \label{sec:data_and_notations}
  
        The data that we use are daily returns of $N=84$ different futures, from $t_\i{beg} = 1/1/2009$ to $t_\i{end} = 1/1/2019$. The list of the different futures is given in Appendix A. None of the chosen assets are exchanged on the Asian futures market, in order to avoid spurious correlations  between returns labelled with different dates, arising from the offsets in the market opening times.

        The futures that we consider are classified in four different sectors: indexes (IDX), commodities (CMD), bonds (YLD) and foreign exchange rates (FXR). In the FXR sector, all currencies are defined with respect to the US Dollar.  
        
        The daily returns at time $t$ are defined as the difference between the prices at time $t$ and $t - 1$, divided by the standard deviation of the price estimated using the last thirty days.\footnote{This allows us to remove part of, but not all, local volatility fluctuations. The normalized returns are furthermore clipped between $-5$ and $5$, to remove very extreme price changes and/or errors.} They are denoted by $r_\i{k}(t)$, where $\i{k}$ indexes a given contract. We also redefine these returns as to have zero-mean and unit standard deviation: 
        \begin{equation}
            {r}_\i{k}(t) \leftarrow \frac{r_\i{k}(t)
            - \avg{r_\i{k}}}{\sigma_\i{k}}
            \label{eq:def_normalized_returns}
        \end{equation}
        where $\avg{r_\i{k}}$ denotes the average return over the time window $T$ and $\sigma_\i{k}$ its standard deviation. 

        We also define the vector of returns ${\mathbf{r}}(t) \coloneqq ({r}_1(t), \ldots, {r}_\i{N}(t))^\intercal \in \mathbb{R}^N$, and the instantaneous correlation matrix 
        \begin{equation}
            \rho_{\i{i, j}}(t) \coloneqq
            {r}_\i{i}(t) {r}_\i{j}(t).
            \label{eq:def_instantaneous_correlation}
        \end{equation}
        We will denote by $\mathbf{R}$ the matrix with ${\mathbf{r}}(t)$ as its $t$-th columns. The futures market index at day $t$ is denoted by $I_0(t)$ and is defined as
	        \begin{equation}
	            I_0(t) \coloneqq \frac{1}{N}
	            \sum\limits_{i = k}^N s_\i{k} {r}_\i{k} (t)
	            \label{eq:def_index}
        \end{equation}
        where
        \begin{equation}
            s_\i{k} = 
            \begin{cases}
                1 &\text{if }  k \in \text{YLD}\\
                -1 &\text{if } k \notin \text{YLD}
            \end{cases}
            \label{eq:def_si}
        \end{equation}
        and YLD denotes the set of futures in the bonds sector. This sign flip anticipates the fact that the top eigenvector of the correlation matrix has a positive sign on IDX, CMD and FXR sectors and a negative sign on YLD.

        For F $ \in \{\text{IDX, YLD, CMD, FXR} \}$, we define the sub-index $I_\i{F}(t)$ as
        \begin{equation}
            I_\i{F}(t) \coloneqq \frac{1}{|\i{F}|}
            \sum\limits_{k \in \i{F}} {r}_\i{k}(t)
            \label{eq:def_subindex}
        \end{equation}
        
        All factors \eqref{eq:def_index} and \eqref{eq:def_subindex} have zero-mean, as the returns are centered. We further scaled them to be of unit variance.
       
        The eigenvalues of an $N \times N$ symmetric matrix $\mathbf{A}$ are ranked in descending order and are denoted by
        \begin{equation}
            \eigval{A}{1} \geq \ldots \geq \eigval{A}{N}
            \label{eq:def_eigenvalues}
        \end{equation}
        The corresponding unit-norm eigenvectors are denoted by $\eigvec{A}{1}, \ldots, \eigvec{A}{N}$.
        
\begin{figure}[ht]
            \centering
            \includegraphics[width=1\textwidth]{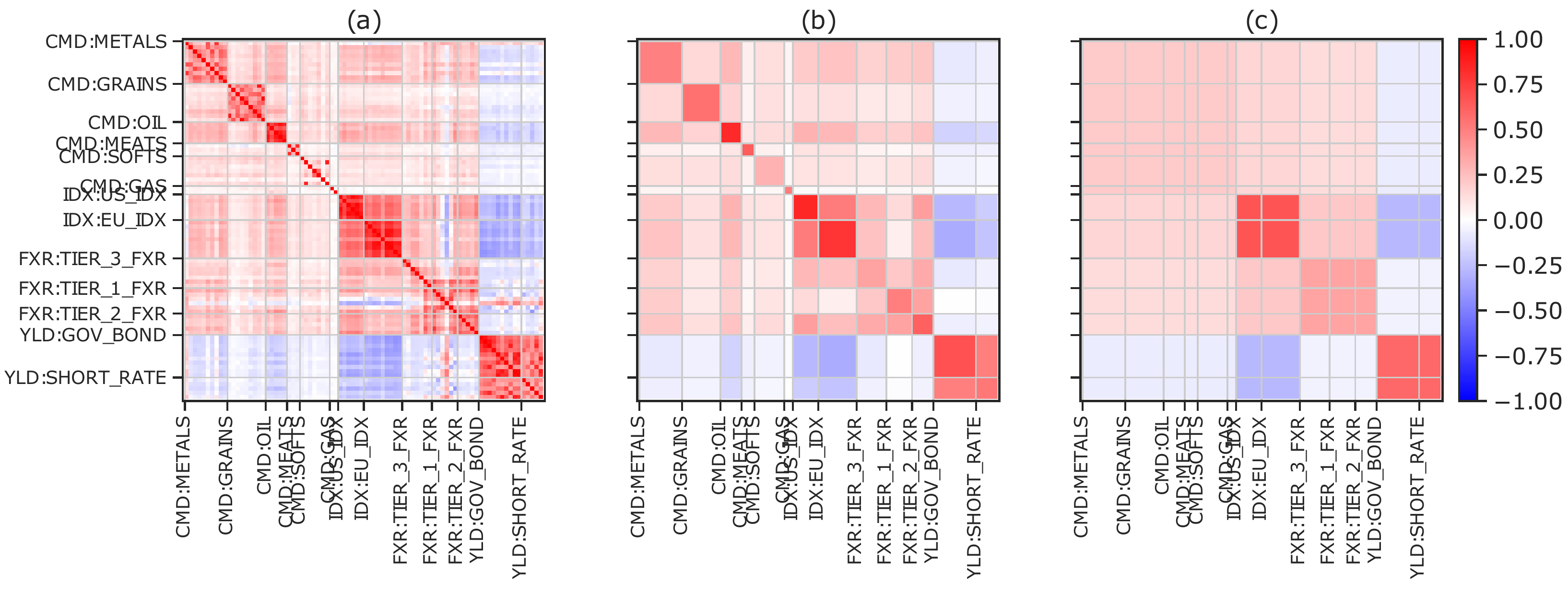}
            \caption{
                (a) Empirical correlation matrix $\Cemp$ for the futures data used in the present paper
                (b) Subsector-averaged version
                (c) Sector-averaged version.
            }
            \label{fig:empirical_correlation_matrix}
        \end{figure}

\begin{figure}[ht]
            \centering
            \includegraphics[width=0.9\textwidth]{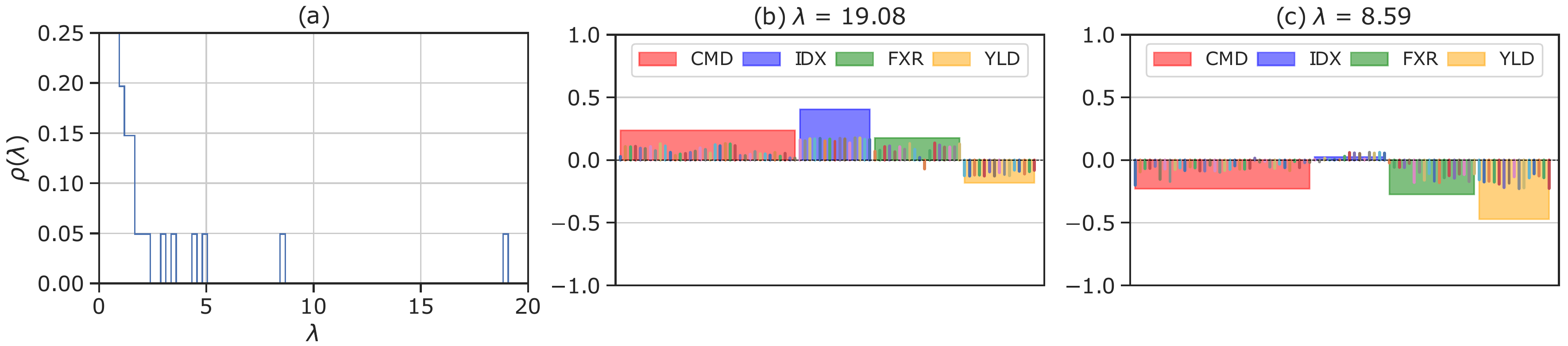}
            \caption{
                (a) Normalized spectrum of the empirical correlation matrix
                (b) \& (c) Visual representation of the two dominant eigenvectors of the empirical correlation matrix titled with their associated eigenvalues.
            }
            \label{fig:empirical_spectrum_eigenvector1_2}
        \end{figure}

        Before turning to our Principal Regression Analysis, we analyze the average correlation matrix $\mathbf{C} = (T - 1)^{-1} \mathbf{R}^\intercal\mathbf{R}$ obtained from the data. It is depicted in Fig.~\ref{fig:empirical_correlation_matrix}.a, sorted by sectors and sub-sectors. In the same figure we also show a version of this matrix with averaged correlations within individual subsectors (Fig.~\ref{fig:empirical_correlation_matrix}.b) and sectors (Fig.~\ref{fig:empirical_correlation_matrix}.c). The spectrum of $\mathbf{C}$ is  shown in Fig.~\ref{fig:empirical_spectrum_eigenvector1_2}.a . The eigenvectors associated to the largest eigenvalue (market mode) and second largest eigenvalue are Visually represented in Fig.~\ref{fig:empirical_spectrum_eigenvector1_2}.b and Fig.~\ref{fig:empirical_spectrum_eigenvector1_2}.c, respectively. We find $\eigval{C}{1} = 19.08$ and $\eigval{C}{2} = 8.59$.
	            
        Let us introduce $\mathbf{e}_0$ the uniform mode for our pool of futures. Given that assets in the YLD sector are anti-correlated with the rest of the pool's assets (see Fig.~\ref{fig:empirical_spectrum_eigenvector1_2}.b), we define it as:
        \begin{equation}
            \mathbf{e}_0 = N^{-1/2}(s_\i{k})_{1 \leq k \leq N}
            \label{eq:def_e0}
        \end{equation}
        where $s_\i{k}$ is defined in \eqref{eq:def_si}. The top eigenvector is such that  $\scal{\mathbf{e}_0}{\eigvec{C}{1}} = 0.90$, meaning that the eigenvector $\eigvec{C}{1}$ is indeed quite similar to the uniform mode.

    \section{Principal Regression Analysis}
    \label{sec:pra}


        \subsection{A Naive Index Factor}
        \label{subsec:global_mean_pred}

\begin{figure}[ht]
                \begin{minipage}{0.475\textwidth}
                    \centering
                    \includegraphics[width=0.75\textwidth]{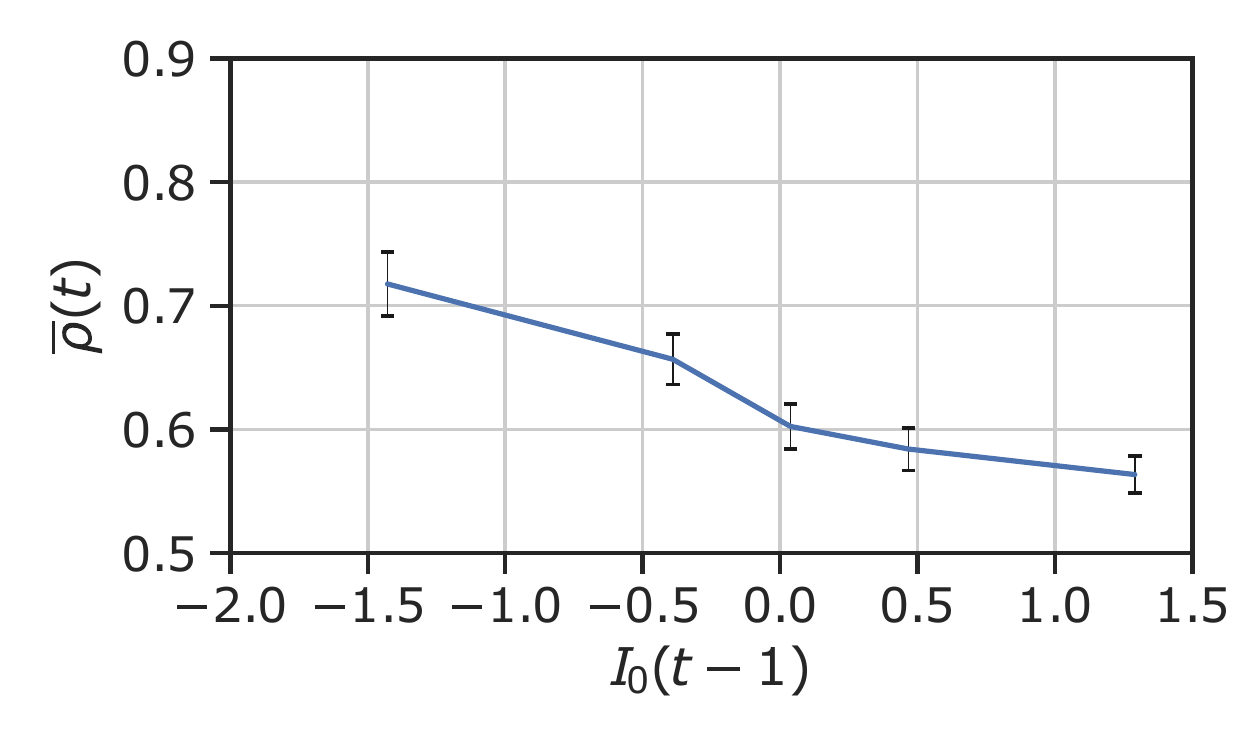}
                    \caption{Average (signed) correlation $\overline{\rho}(t)$, defined in \eqref{rhobar_def}, as a function of the past day index $I_0(t - 1)$. The plot is obtained by binning and averaging values of $\overline{\rho}(t)$ over five ranges of $I_0(t - 1)$ of equal length, $(\i{max}\,I_0(t-1) - \i{min}\,I_0(t-1))/5$.}
                    \label{fig:avg_corrs_vol_I}
                \end{minipage}%
                \hfill
                \begin{minipage}{0.475\textwidth}
                    \centering
                    \includegraphics[width=0.75\textwidth]{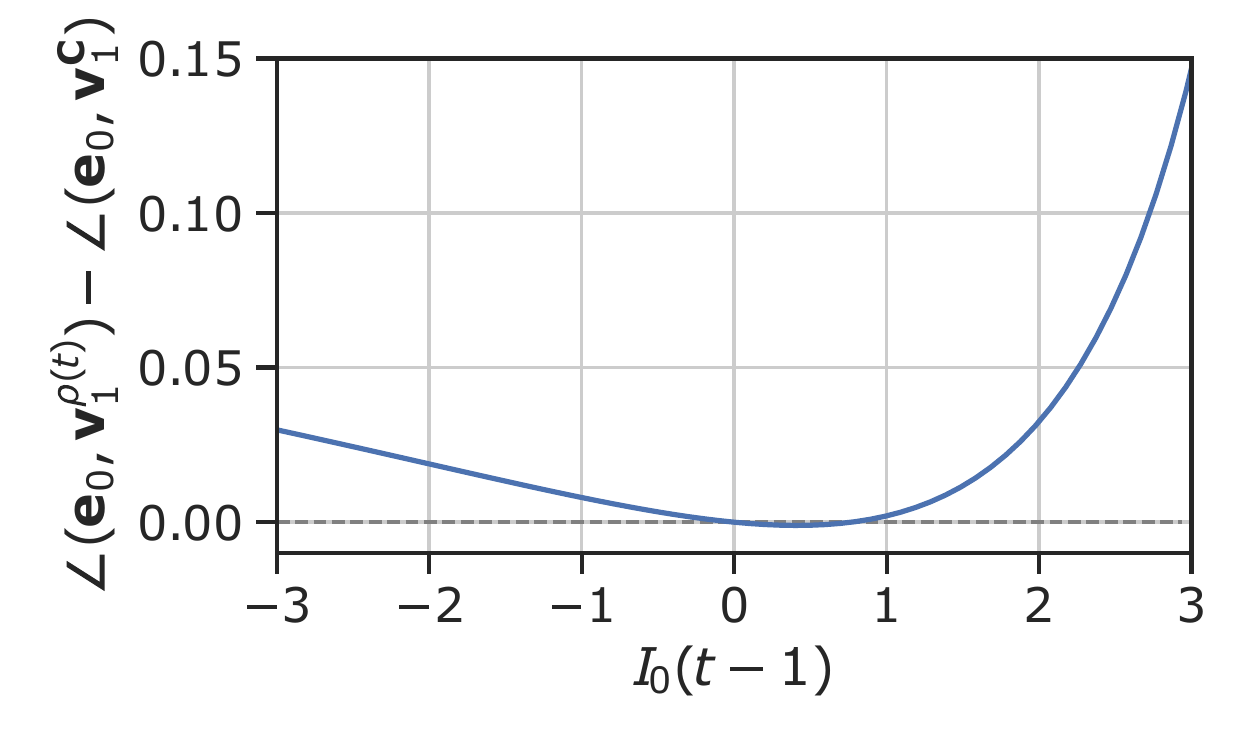}
                    \caption{Angle between the instantaneous, $I_0$ dependent correlation matrix top eigenvector and the uniform mode $\mathbf{e}_0$ compared to the angle between the unconditional correlation matrix top eigenvector $\eigvec{C}{1}$ with $\mathbf{e}_0$. The $\angle$ symbol denotes the $\mathrm{arccos}$ function. We measure $\angle(\mathbf{e}_0,\eigvec{C}{1}) = 0.45\,\mathrm{rad}$.}
                    \label{fig:uniform_mode_dev_simple_index}
                \end{minipage}
            \end{figure}

           We first consider a linear statistical model similar to what is considered in \cite{reigneron2011principal}, only using the futures market sign-adjusted index $I_0$ defined in \eqref{eq:def_index} instead of the mean index for stocks. We therefore consider a statistical model such that for each pair of futures indexed by $i, j$, the instantaneous correlation at time $t$ is a linear function of the index $\tau$ days ago. More precisely, the hypothesized model reads:
            \begin{equation}
                \rho_\i{i, j}(t) = \mathbf{C}_\i{i, j}
                + \Dmat_{\i{i, j}}(\tau) I_0(t - \tau)
                + \varepsilon_\i{i, j}(t, \tau)\, ,
                \label{eq:simple_reg_model}
            \end{equation}
            with $ \rho_\i{i, j}(t):=r_\i{i}(t)r_\i{j}(t)$ and $\tau$ a certain lag that we will take to be equal to one day in the present section. 
            The plot of Fig.~\ref{fig:avg_corrs_vol_I} makes the case for hypothesizing such a statistical model: it shows the value of the average signed correlations, defined as
            \begin{equation}\label{rhobar_def}
              \overline{\rho}(t) = \frac{1}{N(N-1)} \sum_{i \neq j} {s_\mathrm{i}s_\mathrm{j} \rho_\i{i,j}}(t)  
            \end{equation} 
            as a function of the past day index $I_0(t - 1)$. A negative, close to linear relationship is clearly visible. Given that $N \overline{\rho}(t)$ can be written as $\scal{\mathbf{e}_0}{{\bf \rho}(t)}\mathbf{e}_0 \approx \scal{\eigvec{\rho}{1}(t)}{{\bf \rho}(t)}\eigvec{\rho}{1}(t) = \eigval{\rho}{1}(t)$, this motivates the fact that \eqref{eq:simple_reg_model} is a reasonable model, as there is a simple linear relationship between $\eigval{\rho(t)}{1}$ and $\eigval{\mathbf{D}}{N}$ -- see below in \eqref{eq:effect_of_index}.

            \begin{figure}[ht]
                \centering
                \includegraphics[width=1\textwidth]{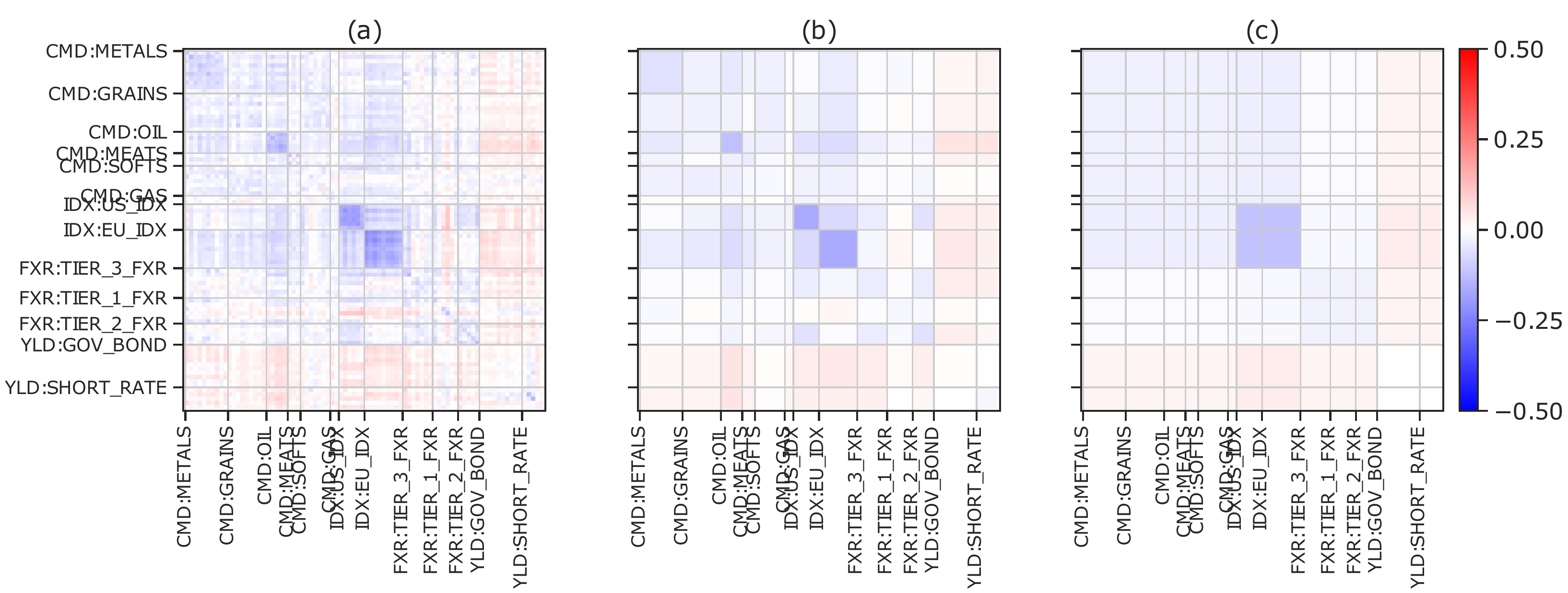}
                \caption{
                    (a) $\Dmat$ matrix obtained by fitting the statistical model \eqref{eq:simple_reg_model}.
                    (b) Subsector-averaged version.
                    (c) Sector-averaged version.
                }
                \label{fig:d}
            \end{figure}

            \begin{figure}[ht]
                \centering
                \includegraphics[width=1\textwidth]{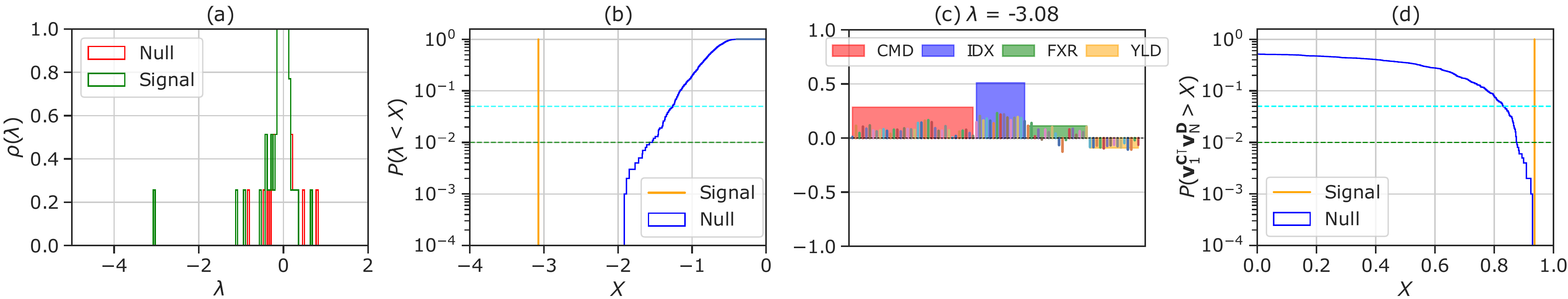}
                \caption{
                    (a) Spectrum of the $\Dmat(\tau=1)$-matrix obtained by fitting the statistical model \eqref{eq:simple_reg_model}, superposed to the spectrum of the null-hypothesis.
                    (b) $\eigval{D}{N}$ and its corresponding CDF in the null-hypothesis case.
                    (c) Visual representation of $\eigvec{D}{N}$.
                    (d) Its overlap with the market mode $\eigvec{\Cemp}{1}$ and the CDF of the corresponding quantity in the null-hypothesis case.
                }
                \label{fig:d_spectrum_cdf_eigenvector1}
            \end{figure}
            
            For a given $\tau$, the slope coefficients $\Dmat_\i{i, j}(\tau)$ are determined by OLS (Ordinary Least-Squares). As alluded to in the introduction,  $\avg{I} = 0$ ensures that the intercept is precisely the $i, j$ entry of the empirical correlation matrix $\mathbf{C}$. 

            The plots in Fig.~\ref{fig:d} depict the obtained $\Dmat(\tau=1)$-matrix (simply denoted by $\Dmat$ in the following), which exhibits a remarkable structure. Its ranked-eigenvalue
            spectrum is depicted in Fig.~\ref{fig:d_spectrum_cdf_eigenvector1}.a, superposed to the null-hypothesis spectrum\footnote{Throughout this paper the null-hypothesis is computed by averaging over 1000 $\Dmat$-matrices, each of which is obtained by fitting the model \eqref{eq:simple_reg_model} to a centered and standardized random predictor, independent of the returns, in place of $I_0(t - 1)$ or $E_\beta(t-1)$.}. We find in particular that the most negative eigenvalue $\eigval{D}{N}$ is highly significant, and equal to  $-3.1 \enskip (p = 0.01)$. All other eigenvalues are not significant.  
            
            As far as eigenvectors are concerned, $\eigvec{D}{N}$ is represented in Fig.~\ref{fig:d_spectrum_cdf_eigenvector1}.c. We find that  $\scal{\eigvec{C}{1}}{\eigvec{D}{N}} = 0.93 \enskip (p = 0.01)$, this quantity being represented along with the CDF of the corresponding null-quantity in Fig.~\ref{fig:d_spectrum_cdf_eigenvector1}.d. Hence, $\eigvec{D}{N}$ is nearly co-linear with the top eigenvector of $\mathbf{C}$. We also find that all other pairs $|\scal{\eigvec{C}{i}}{\eigvec{D}{j}}|$ are much smaller than unity and not significant. 
            
            Hence, we find that the instantaneous correlation matrix $\rho_{ij}(t)$  has all its eigenvectors and eigenvalues essentially independent of $I_0(t-1)$, {\it except for the top one eigenvalue}, which can be written as:
            \begin{equation}
                \eigval{\rho}{1}(t) \approx 
                \eigval{C}{1} + I_0(t-1) \eigval{D}{N} (\scal{\eigvec{C}{1}}{\eigvec{D}{N}})^2 .
                \label{eq:effect_of_index}
            \end{equation}
            In other words, a negative (resp. positive) past index tends to increase (resp. decrease) the largest eigenvalue of the correlation matrix, i.e. increase (resp. decrease) local correlations between futures that are correlated in the market mode and increase (resp. decrease) anti-correlations between anti-correlated ones. Since $\scal{\eigvec{C}{1}}{\eigvec{D}{N}} \approx 1$, the behaviour of $I_0$ hardly changes the direction of the top eigenvector. If anything, the top eigenvector $\eigvec{\rho(t)}{1}$ of the corrected correlation matrix $\rho(t)$ deviates from the uniform mode for negative past-day values of $I_0$, an effect even more pronounced for positive values of this index. These results can be inferred from the plot in Fig~.\ref{fig:uniform_mode_dev_simple_index}. This behavior of the futures market is in contrast with that of stocks presented in \cite{reigneron2011principal}, for which negative index values resulted in the market mode rotating towards the uniform mode in the subsequent days.
            
            We also find that eigenvectors of $\Dmat$ other than $\eigvec{D}{1}$ do not lie in a subspace spanned by a simple subset of $\Cemp$'s eigenvectors, neither are they easily interpretable in terms of sectors. For example, to generate a vector such that its dot product with $\Dmat$'s second eigenvector is greater than $0.9$, we need to include at least five higher-order eigenvectors of $\Cemp$.

            Yet, the matrix plots of Fig.~\ref{fig:d} give some insight about the distribution of the correlations correction across different futures. In particular, it is clear that correlations between assets from the IDX sector tend to increase more than others for a given level of negative past-day index. 


        \subsection{Exponential Moving Average Index}
        \label{subsec:ema_index}

            We now consider a modified index that is weighted by an exponentially-decreasing kernel. More precisely, we define
            \begin{equation}
                J_{\beta}(t) \coloneqq \sum\limits_{s =
                0}^{T_\i{k}} I_0(t - s) \i{e}^{-\beta s}
                \label{eq:index_kernel}
            \end{equation}
            where $T_\i{k}$ is the cutoff time defined as $T_\i{k} = \beta^{-1}\i{log}(100)$, such that $e^{-\beta T_k}=0.01$. Again, we further normalise $J_\beta(t)$ so that it has unit variance. The corresponding $\Dmat_{\beta}$ is then obtained by fitting model \eqref{eq:simple_reg_model}  with $I_0(t-1)$ replaced by $J_\beta(t-1)$. 

            \begin{figure}[ht]
                \centering
                \includegraphics[width=1\textwidth]{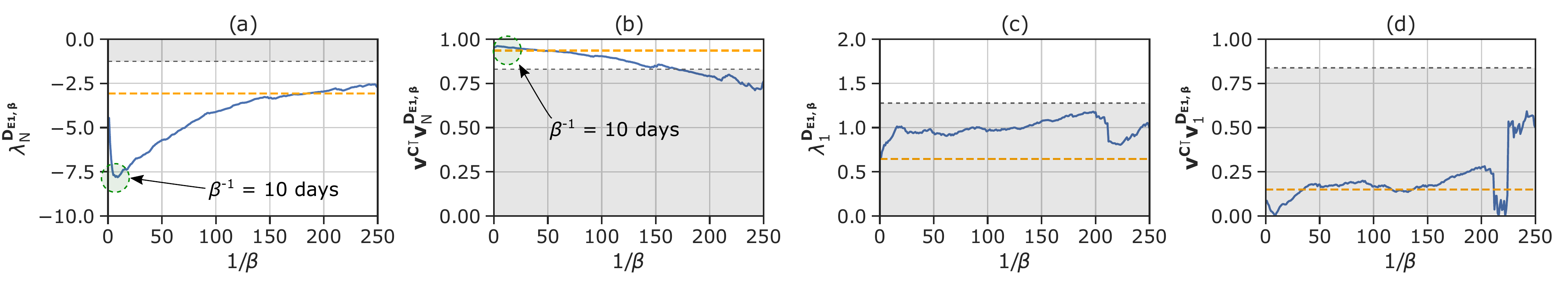}
                \caption{
                    Evolution of some of $\Dmat_\beta$'s spectral characteristics as a function of the decay time
                    $\beta^{-1}$.
                    (a) $\Dmat_\beta$'s most negative eigenvalue;
                    (b) Dot product of associated eigenvector with the market mode $\eigvec{\Cemp}{1}$;
                    (c) $\Dmat_\beta$'s most positive eigenvalue;
                    (d) Dot product of associated eigenvector with the market mode $\eigvec{\Cemp}{1}$.
                    All grey zones indicate non-significant values down to $p = 0.05$ level. Orange dashed lines represent the value of the corresponding quantity for $\Dmat(\tau=1)$.
                }
                \label{fig:d_vs_beta_evolution_simple_index}
            \end{figure}

            Fig.~\ref{fig:d_vs_beta_evolution_simple_index}.a shows the evolution of $\Dmat_\beta$'s most negative eigenvalue against the value of the decay time $\beta^{-1}$, while the alignment of the associated eigenvector with the market mode $\eigvec{\Cemp}{1}$ is depicted in Fig.~\ref{fig:d_vs_beta_evolution_simple_index}.b. From these plots, we can  infer that a timescale of $\beta^{-1} = 10$ days maximizes the magnitude of the effect. The largest \emph{positive} eigenvalue of $\Dmat_\beta$, on the other hand, does not reveal any notable features, see Fig.~\ref{fig:d_vs_beta_evolution_simple_index}.c \& .d. In the following, $\Dmat_\beta$ denotes $\Dmat_{\beta=0.1}$.

            \begin{figure}[t!]
                \centering
                \includegraphics[width=1\textwidth]{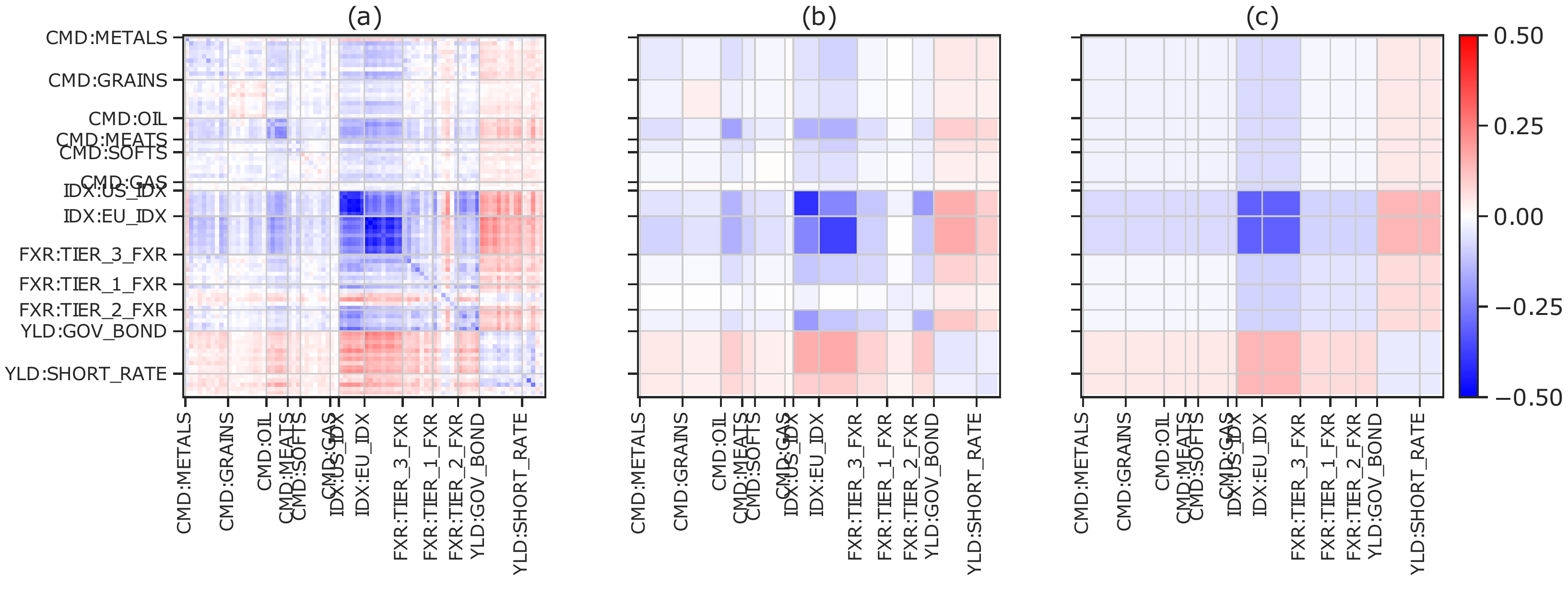}
                \caption{
                    (a) $\Dmat_\beta$ matrix obtained by fitting the statistical model \eqref{eq:simple_reg_model}, with $J_\beta$ in place of $I_0$ and for the optimal timescale parameter $\beta^{-1} = 10$ days.
                    (b) Subsector-averaged version.
                    (c) Sector-averaged version.
                }
                \label{fig:d_hat}
            \end{figure}

            \begin{figure}[t!]
                \centering
                \includegraphics[width=1\textwidth]{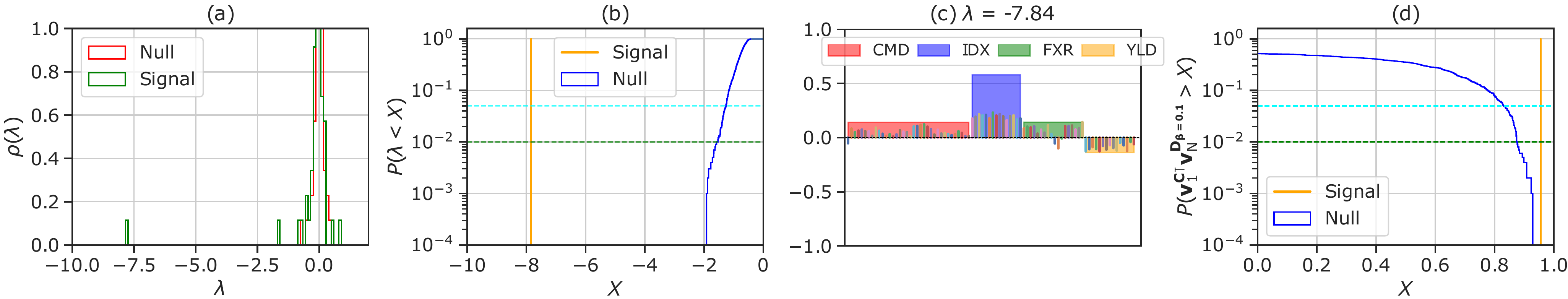}
                \caption{
                    (a) Spectrum of the $\Dmat_\beta$-matrix obtained by fitting the statistical model \eqref{eq:simple_reg_model}, with $J_\beta$ in place of $I_0$, where $\beta=0.1$, the optimal timescale. This spectrum is superposed to that of the null-hypothesis case.
                    (b) $\eigval{D_\beta}{N}$ and its corresponding CDF in the null-hypothesis case.
                    (c) Visual representation of $\eigvec{D_\beta}{N}$.
                    (d) Its overlap with the market mode $\eigvec{\Cemp}{1}$ and the CDF of the corresponding quantity in the null-hypothesis case.
                }
                \label{fig:d_hat_spectrum_cdf_eigenvector1}
            \end{figure} 

            The matrix $\Dmat_{\beta=0.1}$ is depicted in Fig.~\ref{fig:d_hat}, and some important spectrum characteristics are shown in Fig.~\ref{fig:d_hat_spectrum_cdf_eigenvector1}. Both the value of the most negative eigenvalue and of the dot product with the market mode are increased compared to the instantaneous case $\Dmat(\tau=1)$: we measure $\eigval{{D}_\beta}{N} = -7.8 \enskip (p < 0.01)$ and $\scal{\eigvec{C}{1}}{\eigvec{{D}_\beta}{N}} = 0.96 \enskip (p < 0.01)$. As expected, the feedback between past market behaviour and future correlations takes some time to build up; we find that 10 trading days (two weeks) is the order of magnitude of this reaction time.
            
            Note that all other eigenvalues (in particular positive eigenvalues) and dot products are much smaller in absolute value, and therefore not significant, as in Sec.~\ref{subsec:global_mean_pred}.


        \subsection{The ``Eigen-Factor''}
        \label{subsec:eigen-factor_pred}

            \begin{figure}[ht]
                \centering
                \includegraphics[width=1\textwidth]{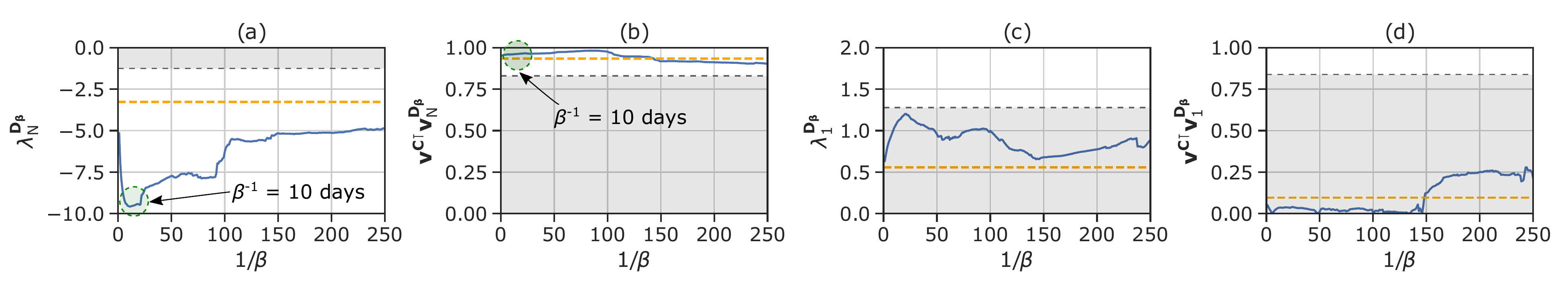}
                \caption{
                    Evolution of some of $\Dmat_{E_\beta}$'s spectral characteristics as a function of the timescale parameter $\beta^{-1}$.
                    (a) $\Dmat_{E_\beta}$'s smallest eigenvalue
                    (b) Dot product of associated eigenvector with the market mode $\eigvec{\Cemp}{1}$
                    (c) $\Dmat_{E_\beta}$'s largest eigenvalue
                    (d) Dot product of associated eigenvectors with the market mode $\eigvec{\Cemp}{1}$.
                    All greyed zones indicate non-significant values to the $p = 0.05$ level. Orange dashed lines represent the value of the associated quantity for $\Dmat_{E}$, the matrix obtained by fitting the statistical model \eqref{eq:model_eigen-factor} with the instantaneous index (obtained by taking $T_\i{k} = 0$ in \eqref{eq:eigen-factor_kernel}).
                }
                \label{fig:d_eigen-factor_vs_beta}
            \end{figure}

            \begin{figure}[ht]
                \centering
                \includegraphics[width=1\textwidth]{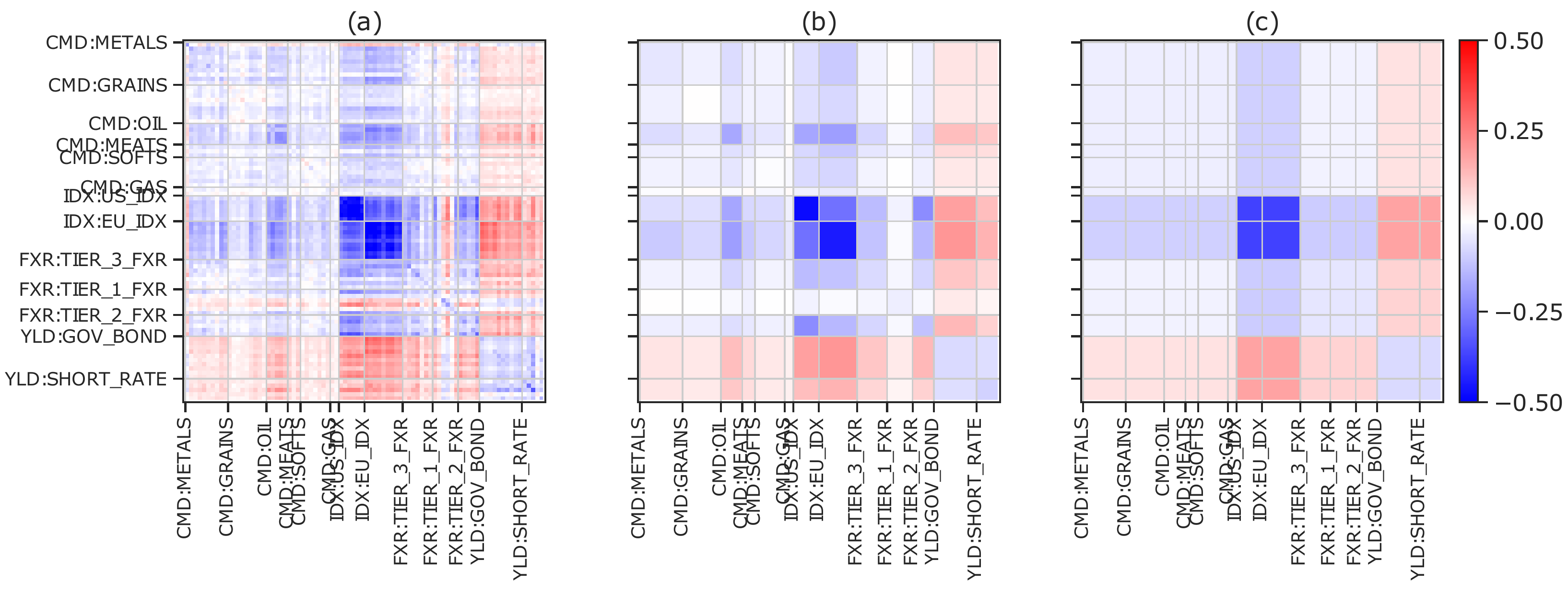}
                \caption{
                    (a) $\Dmat_{E_{\beta}}$ matrix obtained by fitting the
                        statistical model \eqref{eq:model_eigen-factor} with
                        optimal timescale parameter $\beta^{-1} = 10$ days.
                    (b) Subsector-averaged version.
                    (c) Sector-averaged version.
                }
                \label{fig:d_eig1index}
            \end{figure}

            Elaborating on the previous results, we now construct our final indicator, which relies on the local correlation matrix $\Cemp_K(t)$, based on the recent past returns between $t-K$ and $t-1$. 
            
            We chose $K = 3 N$ (days) as a trade-off between finite-size noise and nonstationarity effects. For each $t$, the top eigenvector $\eigvec{C_K}{1}(t)$ of $\Cemp_K(t)$ is determined. The exponentially smoothed ``eigen-factor'' is then computed as:
            \begin{equation}
                E_{\beta}(t) = \sum\limits_{s = 0}^{T_\i{k}} \scal{\mathbf{r}(t)}
                {\eigvec{{\Cemp_K}}{1}}(t)\, \i{e}^{-\beta s}.
                \label{eq:eigen-factor_kernel}
            \end{equation}
            We again make sure that $E_\beta$ is centered and standardized.
            
            We consider the eigen-factor $E_{\beta}(t)$ as a more precise version of $J_\beta(t)$, obtained by fine-tuning each future contract in the index according to its weight in the local market mode while enforcing causality (instead of the $\pm 1$ weights in $I_0(t)$). This procedure does not require ad-hoc changes of signs, like we did for the YLD sector (which might actually change again in the future, as it changed in 1997). Another ``anomaly'' that can be detected in Fig. \ref{fig:empirical_correlation_matrix} is the YEN vs. USD, with a correlation with the IDX sector at odds with other currencies. All these idiosyncracies are automatically accounted for with the eigen-factor $E_\beta(t)$. 
            
            The statistical model for the PRA associated to this eigen-factor reads
            \begin{equation}
                {\bf \rho}(t) = 
                \Cemp +
               \Dmat_{E_\beta}  E_{\beta}(t - 1) +
                {\bf \varepsilon}\, .
                \label{eq:model_eigen-factor}
            \end{equation}
            We first carry out a sensitivity analysis of the dominant eigenvalues of the obtained $\Dmat_{E_\beta}$ matrices against the $\beta$ parameters. The results are depicted in Fig.~\ref{fig:d_eigen-factor_vs_beta}, and advocate for a choice of the optimal timescale parameter $\beta=0.1$, which coincides with that of the $J_\beta$ factor. In the following, $\Dmat_{E}$ will denote $\Dmat_{E_{\beta=0.1}}$. 
            
            \begin{figure}[ht]
                \centering
                \includegraphics[width=1\textwidth]{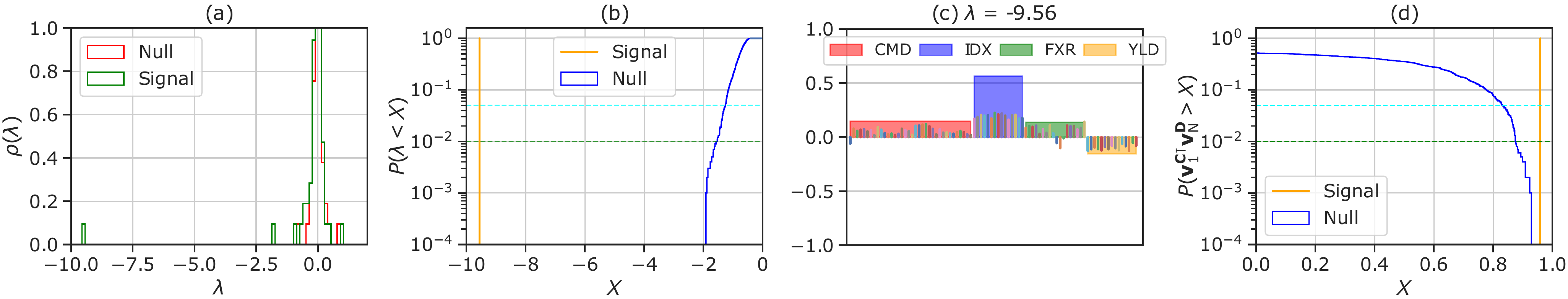}
                \caption{
                    (a) Spectrum of the $\Dmat_{E}$-matrix obtained by fitting the statistical model \eqref{eq:model_eigen-factor}, where $\beta=0.1$, the optimal timescale. This spectrum is superposed to that of the null-hypothesis case.
                    (b) Value of $\eigval{D_{E}}{N}$ and its corresponding CDF in the null-hypothesis case.
                    (c) Visual representation of $\eigvec{D_{E}}{N}$.
                    (d) Its overlap with the market mode $\eigvec{\Cemp}{1}$ and the CDF of the corresponding quantity in the null-hypothesis case.
                }
                \label{fig:d_eig1index_specprops}
            \end{figure}
 
            \begin{figure}[ht]
                \centering
                \includegraphics[width=0.35\textwidth]{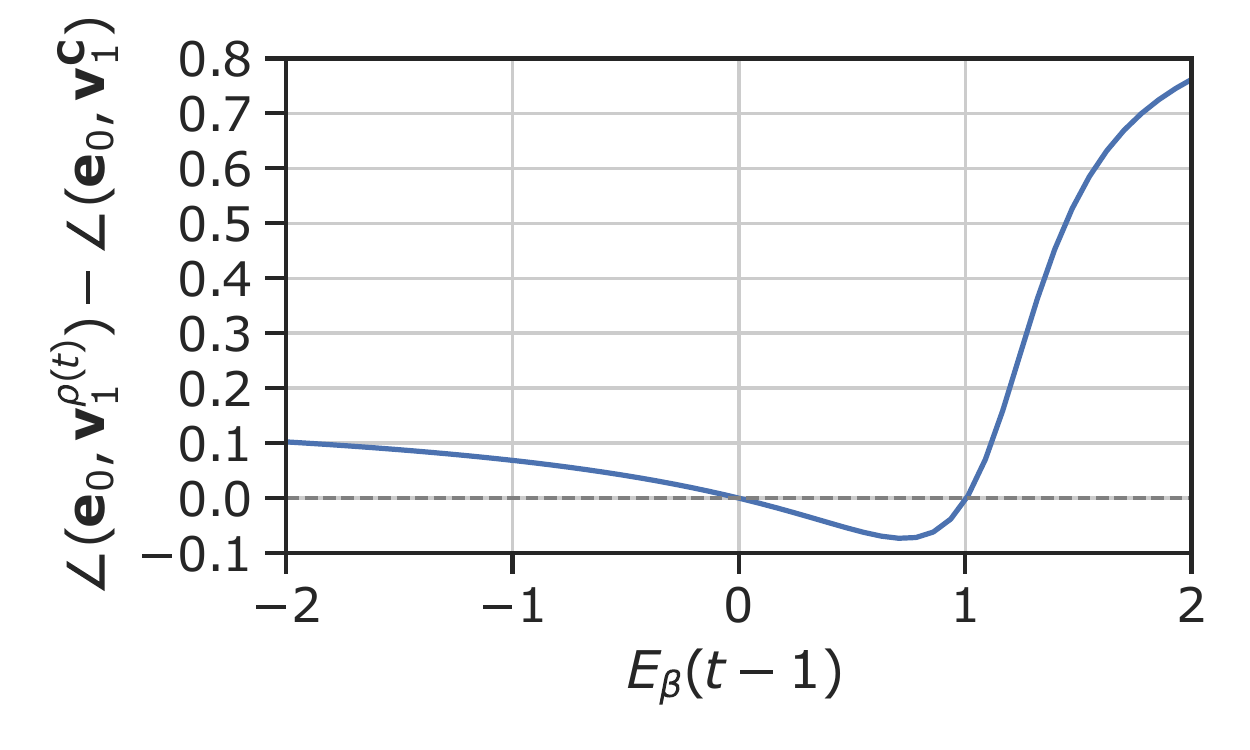}
                \caption{Angle between the instantaneous, $E_\beta$ dependent correlation matrix top eigenvector and the uniform mode $\mathbf{e}_0$ compared to the angle between the unconditional correlation matrix top eigenvector $\eigvec{C}{1}$ with $\mathbf{e}_0$. The $\angle$ symbol denotes the $\mathrm{arccos}$ function. We measure $\angle(\mathbf{e}_0,\eigvec{C}{1}) = 0.45\,\mathrm{rad}$. Note that for moderately positive eigen-factor $E_\beta(t-1) \in [0;1]$, the top eigenvector rotates {\it towards} $\mathbf{e}_0$.}
                \label{fig:uniform_mode_dev_eigen_index}
            \end{figure}

            Let us now describe the results of the PRA using the eigen-factor $E_\beta(t)$ with this optimal timescale parameter.
            The  matrix $\Dmat_{E}$ obtained by fitting the statistical model \eqref{eq:model_eigen-factor} with the optimal timescale parameter $\beta=0.1$ is depicted in Fig.~\ref{fig:d_eig1index}. Note that the structure of $\Dmat_{E}$ appears much more markedly than within our previous two attempts, compare with Fig.~\ref{fig:d} and \ref{fig:d_hat}. The spectral characteristics of $\Dmat_{E}$ are shown in Fig.~\ref{fig:d_eig1index_specprops}. These results indicate that the $E_\beta(t)$ index leads to a significantly stronger effect than what we observed with $J_\beta(t)$. In particular, we measure $\eigval{{D}_{E}}{N} = - 9.56 \enskip (p < 0.01)$ and $\scal{\eigvec{C}{1}}{\eigvec{{D}_{E}}{N}} = 0.96 \enskip (p < 0.01)$, to be compared with, respectively, $-7.8$ and $0.96$ for $J_\beta$.  All other eigenvalues (in particular positive eigenvalues) and dot products are again significantly smaller in absolute value and can be neglected.

            As for the index $I_0$, we measure the difference in deviation of $\rho_{ij}(t)$'s top eigenvector from the uniform mode $\mathbf{e}_0$. The results are depicted in Fig.~\ref{fig:uniform_mode_dev_eigen_index}. The results are qualitatively the same as for the $I_0$ index: when the eigen-factor moves, the top eigenvector of $\rho_{ij}(t)$ moves \emph{away} from the the uniform mode $\mathbf{e}_0$, except for a small region of $E_\beta(t-1) \in [0;1]$ where it rotates towards $\mathbf{e}_0$.
            

        \section{Zooming-in: Sector Indicators}
        \label{subsec:sectorial_predictors}

            In order to break down the effect reported above into different sector contributions,  we now carry out a PRA using corresponding sub-indexes as indicators. More precisely, we consider the mean return of each sector over time scale $\beta^{-1}$, and hypothesize the following statistical model 
            \begin{equation}
                {\bf \rho}(t) = \mathbf{C} +
                \sum\limits_{F \in \mathcal{F}} \Dmat_\i{F} I_\i{F,\beta}(t - 1) 
                + {\bf \varepsilon}\, ,
                \label{eq:multi_reg_model_predef}
            \end{equation}
            with the notations defined in Sec.~\ref{sec:data_and_notations}. Again, we chose the value $\beta^{-1} = 10$ days for the time decay parameter in the following analysis (chosen equal for all sectors).

            \begin{figure}[ht]
                \centering
                \includegraphics[width=1\textwidth]{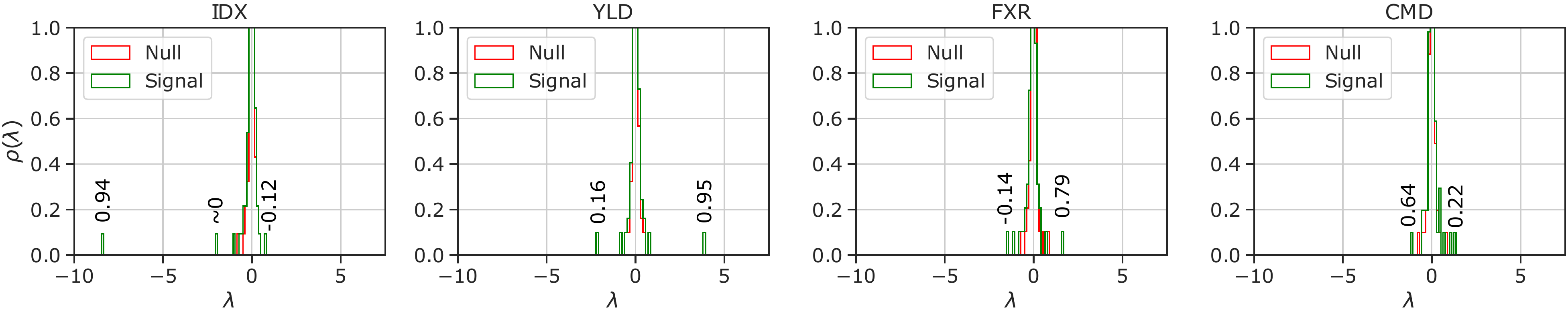}
                \caption{
                    Spectra of the $\Dmat_\i{F}$ matrices obtained by fitting the statistical model \eqref{eq:multi_reg_model_predef}. The numbers reported on top of the extreme eigenvalues give the value of the dot product between the corresponding eigenvector and the market mode $\eigvec{C}{1}$.
                }
                \label{fig:spectrums_sectors}
            \end{figure}

            The spectra of the $\Dmat_\i{F}$ matrices are represented in Fig.~\ref{fig:spectrums_sectors}. The dot products of the eigenvectors corresponding to the extreme eigenvalues with the market mode $\eigvec{C}{1}$ are also reported on the figure. The results show that the only eigenvalues and eigenvectors that carry both value and direction significance are IDX smallest eigenvalue $\eigval{D_\i{IDX}}{N} = -8.5\enskip(p < 0.01)$ and YLD-index largest eigenvalue $\eigval{D_\i{YLD}}{1} = 3.9\enskip(p < 0.01)$. The impact on the correlations following the market mode is then clear through
            \begin{equation}
                \eigval{\rho}{1}(t) \approx 
                \eigval{C}{1}
                + I_\i{IDX, \beta}(t - 1) \eigval{D_\i{IDX}}{N} (\scal{\eigvec{C}{1}}{\eigvec{D_\i{IDX}}{N}})^2 
                + I_\i{YLD, \beta}(t - 1) \eigval{D_\i{YLD}}{1} (\scal{\eigvec{C}{1}}{\eigvec{D_\i{YLD}}{1}})^2,
                \label{eq:effect_of_index_sectorial}
            \end{equation}
            which shows that negative (resp. positive) past IDX-index increases (resp. decreases) the correlations along the market mode, while the opposite holds for the past YLD-index. However, the magnitude of the effect is approximately two times larger for the IDX index compared to YLD, given the relative values of $\eigval{D_\i{IDX}}{N}$ and $\eigval{D_\i{YLD}}{1}$.
            
            Given the low magnitude of their eigenvalues, and the unclear direction of their eigenvectors, the effect of the CMD and FXR sector indexes is more difficult to interpret: their overlap with the market mode $\eigvec{C}{1}$ is small. Therefore, quantifying the effect of these indexes on the correlations $\rho(t)$ in an interpretable manner requires to identify a low-dimensional subspace in which the top eigenvectors of $D_{\i{FXR}}$ or  $D_{\i{CMD}}$ live, in the basis of $\Cemp$'s eigenvectors. But as was the case in Sec.~\ref{subsec:global_mean_pred} for higher modes of the $\Dmat$ matrix, we find that the subspaces in which the corresponding eigenvectors live have no simple interpretations. 

       
    \section{Summary and Conclusion}
    
        This work shows that the correlation matrix describing the co-evolution of futures contract depends significantly on the past return of the "futures market mode" (where stock indexes, commodities and currencies have a $+1$ weight and bonds a $-1$ weight). More precisely, our results suggest that negative past returns of this futures market mode lead to an increased forward average correlation. This effect is even more prominent if one considers the so called "eigen-factor", an average index that finely weights each future contract according to its weight in the empirical correlation's matrix dominant eigenmode. We have also identified that an EMA of both indexes with a characteristic timescale of about $10$ days maximizes its explanatory power of the changes in future average correlations.

        Sectorial indexes give us a finer picture of the futures market's index leverage effect, as our analysis indicates that the mean variations of returns within different sectors have different effects on the future correlations, both in sign and magnitude. Namely, past market movements within the indexes (IDX) sector are negatively correlated with a given day's instantaneous correlations, while the opposite is true for bonds (YLD) sector. As the indexes and bond sectors are anti-correlated in the market mode, the sectorial indexes give us a consistent decomposition of the results that were obtained with simple indexes. Moreover, this decomposition gives us a quantitatively finer picture of each sector's contribution to the average correlations movements, with the indexes sector having twice as much effect on the correlations on average than the bonds sector.
        
        This work calls for subsequent investigations in several directions. Firstly, the predictive power of the linear PRA could be tested, using, e.g., an out-of-sample score of the risk as obtained with a PRA-fitted linear model, such as \eqref{eq:simple_reg_model} or \eqref{eq:multi_reg_model_predef}. This score can then be compared  to that obtained through an uncorrected empirical matrix, to assess whether one would gain in using the PRA models for risk estimation purposes in non-stationary regimes. Secondly, the identification of the various predictors considered in this paper opens the path to using them as features in more complex correlations and risk assessment models (e.g. neural networks). Finally, it would be interesting to decompose the effect exposed in this paper into a local volatility contribution (which our local volatility normalisation does not remove entirely) and a genuine local correlation contribution. Work in this direction is underway \cite{morel}. 
        
    \section*{Acknowledgments}
    
        This research was conducted within the \emph{Econophysics \& Complex Systems} Research Chair, under the aegis of the Fondation du Risque, the Fondation de l'Ecole polytechnique, the Ecole polytechnique and Capital Fund Management. We thank Rudy Morel for very useful discussions.


    \bibliographystyle{unsrt}  
    \bibliography{references}
    

    \appendix
    
    \section{List of futures}

        In this appendix, we provide the list of futures used in this study. We organize these contracts by sector, and by market on which they are traded. This list is reported in Table~\ref{table:futures}.
        \vspace{1\baselineskip}
    

\begin{table}[h!]
\tiny
    \caption{List of futures ordered by sectors and markets}
    \label{table:futures}
\hspace*{-1.7cm}
    \begin{tabular}{|l|l|l|l|l|l|}
        \hline
         \cellcolor[HTML]{EFEFEF} & \textbf{Sector} & \multicolumn{1}{c|}{IDX (indexes)} & \multicolumn{1}{c|}{YLD (bonds)} & \multicolumn{1}{c|}{FXR (currencies)} & \multicolumn{1}{c|}{CMD (commodities)} \\ \hline
        \textbf{Market} & \cellcolor[HTML]{EFEFEF} & \cellcolor[HTML]{EFEFEF} & \cellcolor[HTML]{EFEFEF} & \cellcolor[HTML]{EFEFEF} & \cellcolor[HTML]{EFEFEF} \\ \hline
        Chicago Mercantile Exchange & \cellcolor[HTML]{EFEFEF} & \begin{tabular}[c]{@{}l@{}} E-mini Russell 2000 Index Futures,\\ E-Mini NASDAQ 100 Index Future,\\ E-Mini S\&P's MidCap 400 Index Future,\\ E-Mini S\&P's 500 Index Future
        \end{tabular} & 
        \begin{tabular}[c]{@{}l@{}}   3 Month Eurodollar Future
        \end{tabular} & 
        \begin{tabular}[c]{@{}l@{}}Australian Dollar Currency Future,\\Brazilian Real Currency Future,\\ Canadian Dollar Currency Future,\\Swiss Franc Currency Future,\\ Euro Foreign Exchange Currency Future,\\ British Pound Currency Future, \\ Japanese Yen Currency Future,\\ Mexican Peso Currency Future,\\ New Zealand Dollar Currency Future,\\ South African Rand Currency Future
        \end{tabular} & 
        \begin{tabular}[c]{@{}l@{}} Feeder Cattle Future,\\  Live Cattle Future,\\ Lean Hogs Future        \end{tabular} \\
        \hline
        Chicago Board of Trade & \cellcolor[HTML]{EFEFEF} & \begin{tabular}[c]{@{}l@{}}Mini Dow Jones Industrial\\ Average e-CBOT Future
        \end{tabular} & 
        \begin{tabular}[c]{@{}l@{}}10 Year US Treasury Note Future,\\ 2 Year US Treasury Note Future,\\ 5 Year US Treasury Note Future,\\ US Long Bond Future
        \end{tabular} & 
        & 
        \begin{tabular}[c]{@{}l@{}} Corn Future,\\ Hard Red Winter Wheat Future,\\Soybean Meal Future, \\Soybean Oil Future,\\Soybean Future,\\Wheat Future
        \end{tabular} \\ 
        \hline
        New York Mercantile Exchange & \cellcolor[HTML]{EFEFEF} & &  & 
        & 
        \begin{tabular}[c]{@{}l@{}} Light Sweet Crude Oil Future,\\ NY Harbor ULSD Futures,\\  Henry Hub Natural Gas Futures,\\ Palladium Future,\\Platinum Future,\\ Reformulated Gasoline Blendstock\\for Oxygen Blending RBOB Futures, 
        \end{tabular} \\ 
        \hline
        Commodity Exchange, Inc. & \cellcolor[HTML]{EFEFEF} &  &  &   & 
        \begin{tabular}[c]{@{}l@{}}Copper Future, \\ Gold 100 Troy Ounces Future, \\ Silver Future,
        \end{tabular} \\ 
        \hline
        Eurex & \cellcolor[HTML]{EFEFEF} & 
        \begin{tabular}[c]{@{}l@{}}DAX Index Future,\\ EURO STOXX 50 Future,\\ Swiss Market New Index Future
        \end{tabular} & 
        \begin{tabular}[c]{@{}l@{}}5 Year Euro BOBL Future,\\ 10 Year Euro BUND Future,\\ 30 Year Euro BUXL Future,\\ 2 Year Euro SCHATZ Future
        \end{tabular} &  &  \\ 
        \hline
        Euronext Derivatives Paris & \cellcolor[HTML]{EFEFEF} & \begin{tabular}[c]{@{}l@{}} Amsterdam Index Future,\\CAC 40 Index Future
        \end{tabular} &  &  & 
        \begin{tabular}[c]{@{}l@{}}Rapeseed Future,\\Milling Wheat Future
        \end{tabular} \\ 
        \hline
        ICE Futures & \cellcolor[HTML]{EFEFEF} & 
        LSE 100 Index Future &
        \begin{tabular}[c]{@{}l@{}}3 Month Euro Euribor Future,\\ 3 Month Euroswiss Future,\\ Long Gilt Future,\\ 90 Day Sterling Future
        \end{tabular} &  & 
        \begin{tabular}[c]{@{}l@{}}Brent Crude Oil Future,\\Canola Future,\\ Cocoa Future, Gas Oil Future,\\Robusta Coffee Future 10-Tonne,\\White Sugar Future,\\Natural Gas Future, C Coffee Future,\\Number 2 Cotton Future,\\Number 11 World Sugar Future,\\Frozen Concentrated\\Orange Juice A Future,\\
        \end{tabular} \\ 
        \hline
        London Metal Exchange & \cellcolor[HTML]{EFEFEF} &  &  &  & 
        \begin{tabular}[c]{@{}l@{}}Aluminium,\\Lead\\Nickel,\\Tin,\\Zinc
        \end{tabular} \\ 
        \hline
        Montreal Exchange & \cellcolor[HTML]{EFEFEF} & 
        S\&P/TSX 60 Index Future & 
        \begin{tabular}[c]{@{}l@{}}10 Year Canadian Bond Future,\\ 3 Month Canadian Bank Acceptance Future
        \end{tabular} &  &  \\ 
        \hline
        Meff Renta Variable (Madrid) & \cellcolor[HTML]{EFEFEF} & 
        Madrid IBEX 35 Index Future &  &  &  \\ 
        \hline
        Borsa Italiana (IDEM) & \cellcolor[HTML]{EFEFEF} & 
        MIB Index Future &  &  &  \\ 
        \hline
        South African Futures Exchange & \cellcolor[HTML]{EFEFEF} & 
        JSE Top 40 Index Future &  &  &  \\ 
        \hline
        Forward market & \cellcolor[HTML]{EFEFEF} &  &  & 
        \begin{tabular}[c]{@{}l@{}} Czech Koruna,\\Hungarian Forint,\\Israeli New Shekel,\\Norwegian Krone,\\Poland Zloty,\\Swedish Krona,\\Singapore Dollar,\\Turkish lira
        \end{tabular} & \\ 

        \hline
    \end{tabular}
\hspace*{-1.7cm}
\end{table}


\end{document}